\nonstopmode

\newcommand{\ie}{i.e.{}}
\newcommand{\eg}{e.g.{}}
\newcommand{\U}[1]{\,{\rm{#1}}}
\newcommand{\euler}{\textrm{e}}

\newcommand{\HF}{(HF)$_{\infty}$}
\newcommand{\HCl}{(HCl)$_{\infty}$}
\newcommand{\Hartree}{\U{E_{\mathit h}}}

\documentclass{elsart}
\usepackage{graphicx,subeqnarray,cite}
\journal{Chemical Physics Letters}

\begin{document}
\begin{frontmatter}
\title{Basis set convergence in extended systems: infinite hydrogen fluoride
and hydrogen chloride chains}
\author{Christian Buth\corauthref{cor}},
\corauth[cor]{Corresponding author.}
\ead{Christian.Buth@web.de}
\author{Beate Paulus}
\address{Max-Planck-Institut f\"ur Physik komplexer Systeme,
N\"othnitzer Stra\ss{}e~38, 01187~Dresden, Germany}

\begin{abstract}
Basis set convergence of the Hartree-Fock and the correlation energy is
examined for the hydrogen bonded \emph{infinite} bent chains~\HF{} and \HCl{}.
We employ series of correlation consistent basis sets up to
quintuple~$\zeta$ quality together with a coupled cluster method~(CCSD)
to describe electron correlation on \emph{ab initio} level. The
Hartree-Fock energy converges rapidly with increasing basis set quality
whereas the correlation energy is found to be slowly convergent for the
same series of basis sets. We study basis set extrapolation for~\HF{}
and \HCl{} and show that it substantially enhances the accuracy
of both the Hartree-Fock and the correlation energy in
\emph{extended} systems.
\end{abstract}

\end{frontmatter}

\section{Introduction}

\emph{Ab initio} methods for polymers and crystals come more and
more into focus of quantum chemists
and solid state physicists~\cite{Fulde:EC-95,Fulde:WF-02}. As most
\emph{ab initio} treatments of such extended systems rely on incomplete
one-particle basis sets, we consider it timely to investigate the
basis set convergence of Hartree-Fock and correlation energies in
\emph{periodic} systems. \emph{Molecular} Hartree-Fock energies
are well known to converge exponentially, \eg~Ref.~\cite{Jensen:BS-99},
towards the basis set limit, but molecular correlation energies
turn out to converge only with the third power of the highest angular
momentum employed in the basis sets~\cite{Hill:RC-85,Klopper:R12-00}.

The convergence properties of the Hartree-Fock and the correlation
energy can be exploited to extrapolate Hartree-Fock~\cite{Jensen:BS-99} and
correlation energies~\cite{Helgaker:BS-97,Halkier:BS-98,Halkier:BS-99,Klopper:R12-00,Park:BS-01,Huh:BS-03} towards the basis set limit. As only standard methods of
quantum chemistry are required, basis set extrapolation of
correlation energies provides an interesting
alternative over the specialised, explicitly correlated~(R12)
methods, which directly yield near basis set limit wave functions
and correlation energies but have a high computational
demand~\cite{Klopper:R12-00}. Especially
well suited in conjunction with extrapolation schemes, are
the correlation consistent basis sets~\cite{basislib-04}
cc-pVXZ~\cite{Dunning:GBS-89,Woon:GBS-93}, aug-cc-pVXZ~\cite{Dunning:GBS-89,Kendall:EA-92,Woon:GBS-93}
and d-aug-cc-pVXZ~\cite{Dunning:GBS-89,Kendall:EA-92,Woon:GBS-94},
X = D, T, Q, 5, 6 which are hierarchical series of basis sets
of increasing quality.

Our study elucidates the performance of basis set extrapolation
for Hartree-Fock and correlation energies in \emph{infinite} periodic
systems, the hydrogen bonded bent chains~\HF{} and \HCl{}
which are representatives for strong and weak hydrogen
bonds~\cite{Buth:BE-04} and require a very accurate description
by a large one particle basis to reliably determine their binding
energies per monomer. Hartree-Fock energies of the infinite chains are
obtained by periodic calculations~\cite{crystal03} whereas their
correlation energy is calculated utilising Stoll's incremental
scheme~\cite{Stoll:CS-92,Stoll:CD-92,Stoll:CG-92} which has been
successfully applied to various
semiconductors~\cite{Paulus:CE-96,Albrecht:CA-97},
ionic crystals~\cite{Doll:CE-95,Doll:CS-98},
rare gas crystals~\cite{Rosciszewski:AI-99} and
polymers~\cite{Ming:IA-97,Abdurahman:AI-99,Willnauer:QC-04}.

\section{Theory}
\label{sec:theory}

Hartree-Fock energies turn out to converge rapidly with increasing
basis set quality towards the basis set
limit. However, the actual convergence behaviour has only
empirically been determined, Ref.~\cite{Jensen:BS-99} (and
Refs.\ therein), to depend both on the number of basis functions
and on the highest angular momentum in basis sets.
The cardinal number~$X$ of correlation consistent basis sets
is related to both quantities, and Hartree-Fock energies
follow
\begin{equation}
  \label{eq:rhfconv}
  E_{\rm SCF}(\infty) = E_{\rm SCF}(X) - A \, \euler^{-B \, X}  \; ,
\end{equation}
with~$A, B > 0$ and $E_{\rm SCF}(\infty)$~being the Hartree-Fock
basis set limit while the Hartree-Fock energy obtained with a
basis set~$X$ is denoted by~$E_{\rm SCF}(X)$.

Correlation energies converge differently; the partial wave analysis of
the correlation energy of the helium atom~\cite{Hill:RC-85} facilitates
to derive the relation~\cite{Helgaker:BS-97,Halkier:BS-98}
\begin{equation}
  \label{eq:atomicX3series}
  \label{eq:atomicX3}
  E_{\rm corr}(\infty) = E_{\rm corr}(X) - A^{\prime} \, X^{-3} \; ,
\end{equation}
where $E_{\rm corr}(\infty)$~is the basis set limit correlation
energy and $E_{\rm corr}(X)$~represents the correlation
energy obtained with basis set~$X$ (in our case, $X$~is equal to
the highest angular momentum of basis functions in the basis set).

Eq.~(\ref{eq:atomicX3series}) is derived for the asymptotic behaviour,
\ie~large~$X$, of the correlation energy, assuming basis sets of highest
angular momentum~$X$, being centred around a single point in space.
The basis sets are supposed to be complete for all angular
momenta~$\leq X$ and are required to be complete with
respect to their radial part~\cite{Hill:RC-85}.
However, a simple two-point fit based on Eq.~(\ref{eq:atomicX3series}),
which involves the correlation energies of two basis sets~$X$ and $Y$,
turns out to yield highly accurate molecular binding
energies~\cite{Helgaker:BS-97,Halkier:BS-98,Halkier:BS-99}.

The extrapolation scheme for correlation energies of Park, Huh and
Lee~\cite{Park:BS-01,Huh:BS-03} is a more flexible basis set extrapolation
which we consider to cope slightly better with the increasing radial
and angular completeness of hierarchical basis set series. Park~\etal{}
harness
\begin{subeqnarray}
  \label{eq:atomicXgamma}
  E_{\rm corr}^{\rm chain}(\infty) &=& \frac{E_{\rm corr}^{\rm chain}(X)
    - \gamma_{X,Y} E_{\rm corr}^{\rm chain}(Y)}{1 - \gamma_{X,Y}} \\
  \slabel{eq:gammaXX}
  \gamma_{X,Y} &=& \frac{E_{\rm corr}^{\rm monomer}(X  )
    - E_{\rm corr}^{\rm monomer}(\infty)}{E_{\rm corr}^{\rm monomer}(Y)
    - E_{\rm corr}^{\rm monomer}(\infty)} \; ,
\end{subeqnarray}
with the underlying assumption that the basis set convergence
rate~$\gamma_{X,Y}$ is the same for a monomer and an infinite chain
formed by many monomers. $\gamma_{X,Y}$ is the ratio of the absolute
error in the correlation energy of the monomer described by
two different basis sets~$X$ and $Y$. If
the electronic structure of a monomer does not change substantially
upon chain formation, a given basis set represents both the
monomer and the infinite chain equally well.

\begin{figure}
  \includegraphics[clip,width=\hsize]{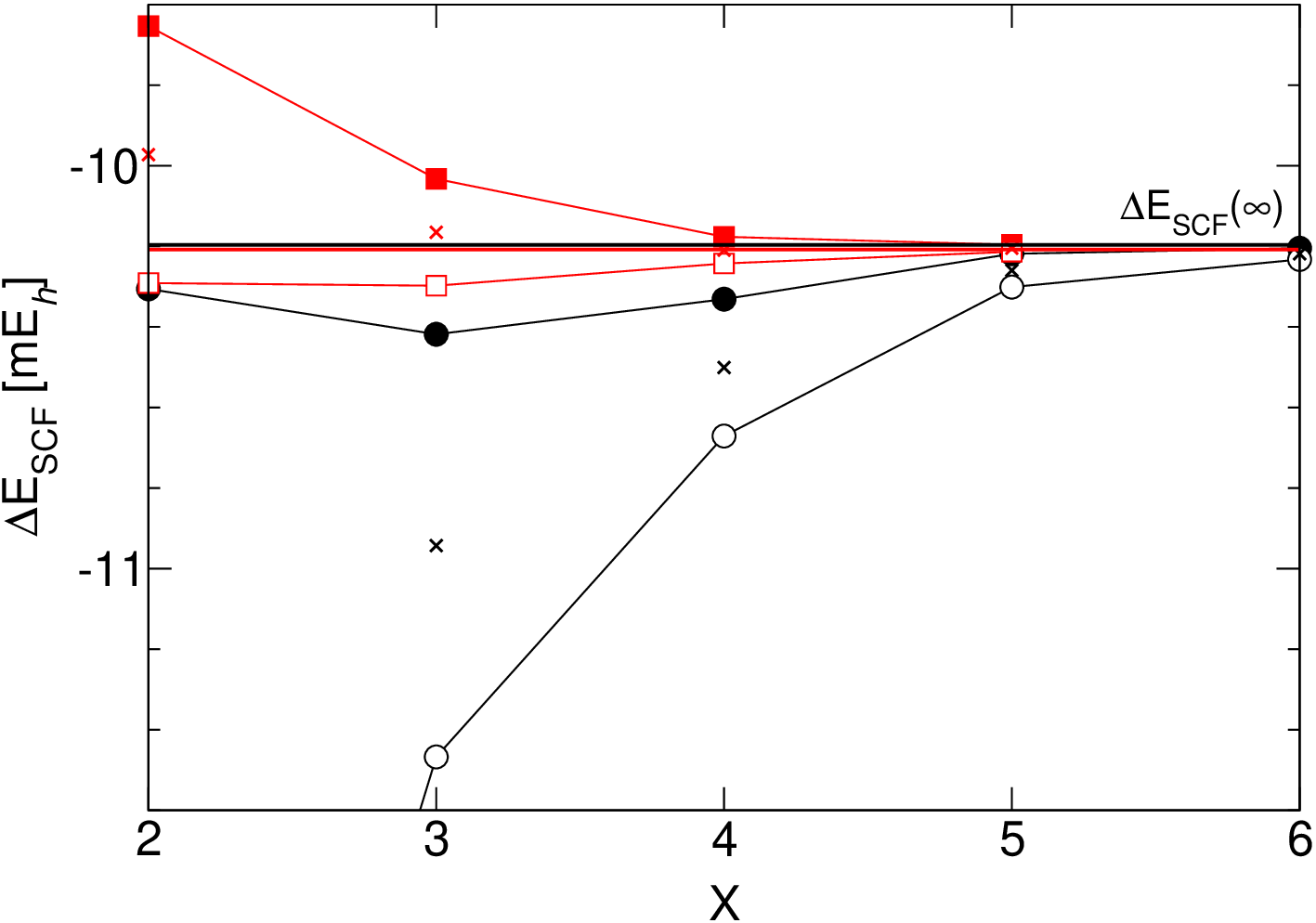}
  \caption{Basis set convergence of the Hartree-Fock binding energy per
           monomer~$\Delta E_{\rm SCF}$ in~\HF. Circles and squares
           represent~$\Delta E_{\rm SCF}(X)$ of the cc-pVXZ and
           aug-cc-pVXZ basis sets where open and closed symbols
           denote bare and CP~corrected Hartree-Fock binding energies.
           The straight line results from two nearly coinciding
           lines which indicate the extrapolated Hartree-Fock binding
           energies, the upper and the lower line referring to the cc-pVXZ
           and the aug-cc-pVXZ basis sets.
           The crosses indicate the mean of the CP~corrected
           and the corresponding bare Hartree-Fock binding energies.}
  \label{fig:HF_RHF_basis}
\end{figure}

\section{Results and discussion}
\label{sec:results}

Basis set extrapolation of Hartree-Fock and correlation energies
shall now be used to obtain accurate binding
energies of \HF\ and \HCl~chains. Both~\HF\ and \HCl~form zig-zag
chains where in both cases the unit cell consists of two monomers.
Details concerning the employed experimental geometries can be found in
Refs.~\cite{Sandor:CS-67,Holleman:IC-01,Buth:BE-04}.

\begin{figure}
  \includegraphics[clip,width=\hsize]{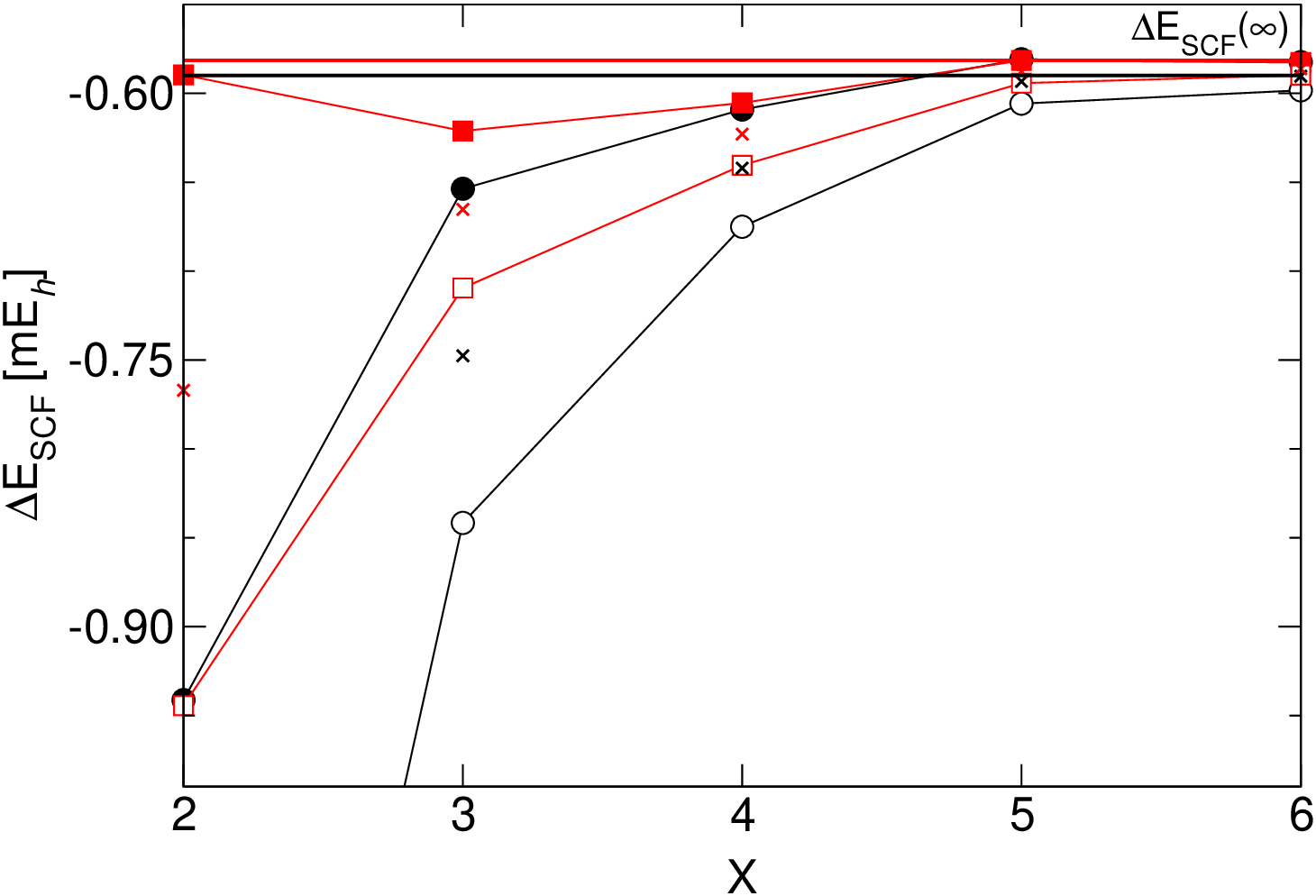}
  \caption{Basis set convergence of the Hartree-Fock binding
           energy per monomer~$\Delta E_{\rm SCF}$ in~\HCl.
           Symbols are chosen as in Fig.~\ref{fig:HF_RHF_basis}.
           The upper and the lower straight lines now refer to
           the aug-cc-pVXZ and the cc-pVXZ basis sets, in
           reverse order compared to Fig.~\ref{fig:HF_RHF_basis}.}
  \label{fig:HCl_RHF_basis}
\end{figure}

The Hartree-Fock binding energies per monomer\footnote{The
Hartree-Fock energies of the infinite chains and the corresponding
monomers were determined with \textsc{crystal}~\cite{crystal03}
where angular momenta, larger than~$s$, $p$, $d$ had to be
removed from the basis sets.} are plotted for~\HF{} and \HCl{}
in Figs.~\ref{fig:HF_RHF_basis} and~\ref{fig:HCl_RHF_basis}.
The basis set superposition error~(BSSE) is removed beyond microhartree
accuracy by counterpoise correction~(CP)~\cite{Boys:CP-70,Duijneveldt:CP-94},
where a monomer is additionally surrounded by the basis functions of
eight neighbouring monomers.
For each series of correlation consistent basis sets, there is an
upper curve for the CP corrected binding energies and a
corresponding lower curve giving the bare binding
energies without CP~correction of the monomer energies. Both curves
converge unsystematically towards the Hartree-Fock basis set limit,
especially, they generally converge not monotonic.

The deviation of the lower curve from the upper curve of the same
basis set series yields an estimate of the error of the approximation
introduced by the finite basis sets as this deviation is the size of the
BSSE~\cite{Duijneveldt:CP-94} and an estimate of the incompleteness
of a one-particle basis set. It is very small,
%
%
%
%
$0.27\%$ for~\HF{} and $2.7\%$ for~\HCl{}, utilising
cc-pV6Z basis sets. Nevertheless, we would like to elucidate whether,
in case of an infinite chain, the Hartree-Fock energies follow
Eq.~(\ref{eq:rhfconv}) as well, \ie~whether the packing in infinite
periodic systems has an unexpected impact on Hartree-Fock basis set
convergence. Tab.~\ref{tab:RHFextra} gives Hartree-Fock
energies at the basis set limit for the monomers and the
infinite chains as obtained by a least
squares fit using Eq.~(\ref{eq:rhfconv}) which turns
out to describe the Hartree-Fock energies, underlying
Figs.~\ref{fig:HF_RHF_basis} and \ref{fig:HCl_RHF_basis}, excellently.
A three-point fit based on Eq.~(\ref{eq:rhfconv}) to the Hartree-Fock
energies, obtained with three basis sets~$X$, $Y$ and $Z$, also
yields convincing results that converge rapidly with
the quality of the three basis sets used.
Halkier~\etal~\cite{Halkier:BS-99} found for the basis set convergence
of the Hartree-Fock binding energy of several
hydrogen-bonded complexes, including (HF)$_2$ and (HCl)$_2$, that
the mean of the bare and the counterpoise corrected Hartree-Fock binding
energies frequently provides a decent extrapolation to the basis set
limit. This behaviour of the mean Hartree-Fock binding energy is
solely observed for the aug-cc-pVXZ series for~\HF{}.

\begin{table}
  \caption{Basis set extrapolated Hartree-Fock binding energies per
           monomer~$\Delta E_{\rm SCF}(\infty)$ of~\HF{} and \HCl{}
           obtained by least squares fits to Eq.~(\ref{eq:rhfconv})
           of their Hartree-Fock energies for the cc-pVXZ ($X={}$D, \ldots, 6) and
           aug-cc-pVXZ ($X={}$D, \ldots, 5 for \HF{} and  $X={}$D, \ldots, 6 for~\HCl{})
           series of basis sets. All data are given in millihartree.}
  \centering
  \begin{tabular}{rccc}
    \hline
      & BSSE & \multicolumn{2}{c}{$\displaystyle \Delta E_{\rm SCF}
                (\infty)$} \\[-1.5ex]
      \raisebox{1.5ex}[-1.5ex]{Compound} & correction & cc-pVXZ & aug-cc-pVXZ \\
    \hline
     \HF  & Non  & -10.199 & -10.194 \\
          & CP   & -10.194 & -10.222 \\
     \HCl & Non  & -0.596 & -0.586 \\
          & CP   & -0.584 & -0.577 \\
    \hline
  \end{tabular}
  \label{tab:RHFextra}
\end{table}

For a systematic treatment of electron correlation in~\HF{} and
\HCl, we expand the correlation energy per monomer in terms of
Stoll's incremental scheme~\cite{Stoll:CS-92,Stoll:CD-92,Stoll:CG-92}.
This yields a decomposition of the correlation energy per monomer
in terms of the correlation energy arising from the electrons of a single
monomer within the infinite chains and non-additive contributions from
electrons of pairs, triples, \ldots{} of monomers.
The individual energy increments are obtained using CCSD (as
implemented in the \textsc{molpro} program~\cite{MOLPRO2002.6,Hampel:CE-92})
to correlate the valence electrons of selected monomers within the
oligomers. The oligomers comprise six to ten monomers and are arranged
in the geometry of the respective chains, for details see
Ref.~\cite{Buth:BE-04}. BSSE is removed by applying the
CP~correction~\cite{Boys:CP-70,Duijneveldt:CP-94}
where only the basis functions of monomers neighbouring the central
monomer of those oligomers, used in the calculation
of the energy increments, are retained.

\begin{figure}
  \includegraphics[clip,width=\hsize]{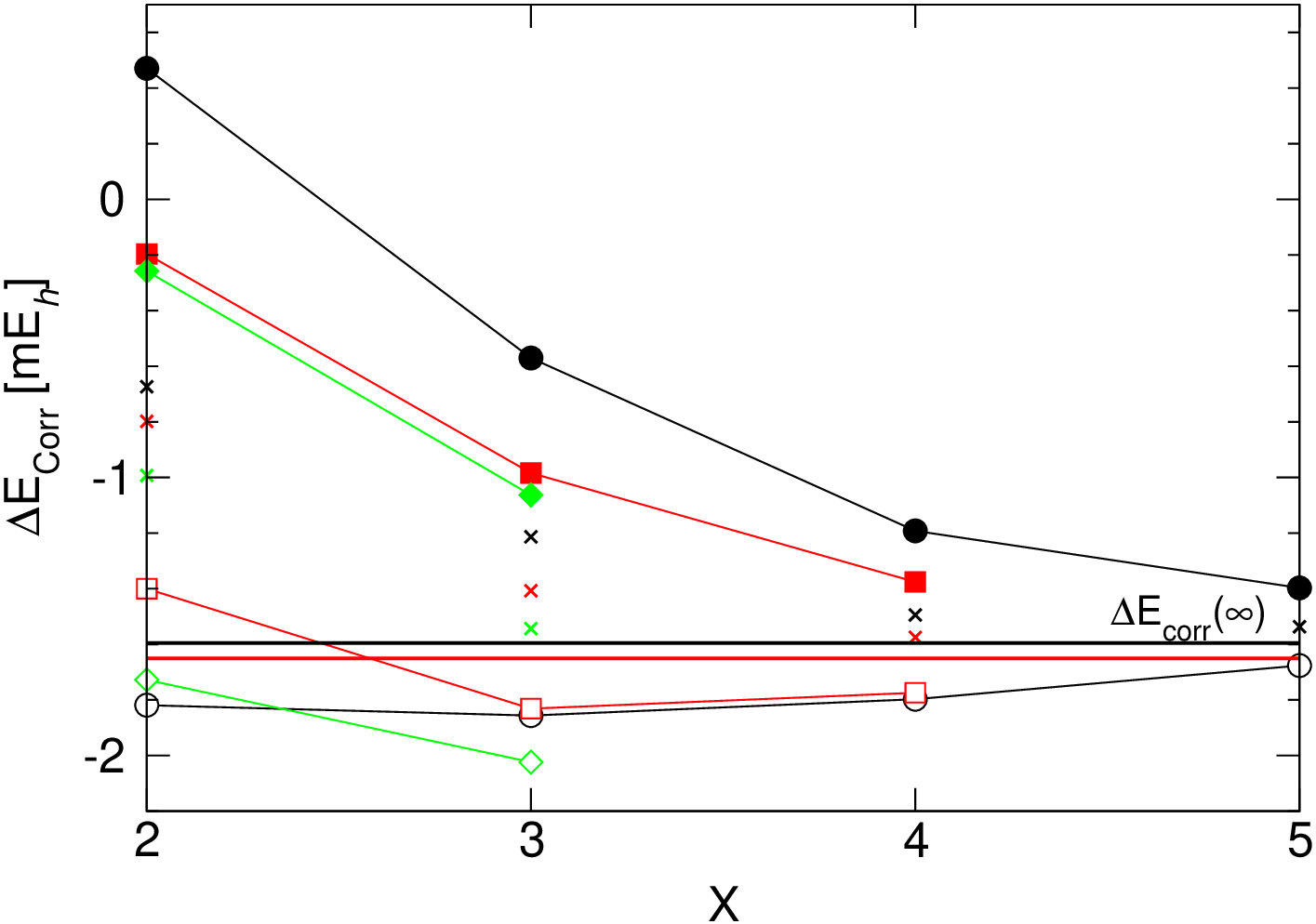}
  \caption{Basis set convergence of the correlation contribution to the
           binding energy per monomer~$\Delta E_{\rm corr}$ in~\HF{}.
           Symbols are chosen as in Fig.~\ref{fig:HF_RHF_basis},
           with diamonds showing data for the d-aug-cc-pVXZ basis sets.}
  \label{fig:HF_basis}
\end{figure}

\begin{figure}
  \includegraphics[clip,width=\hsize]{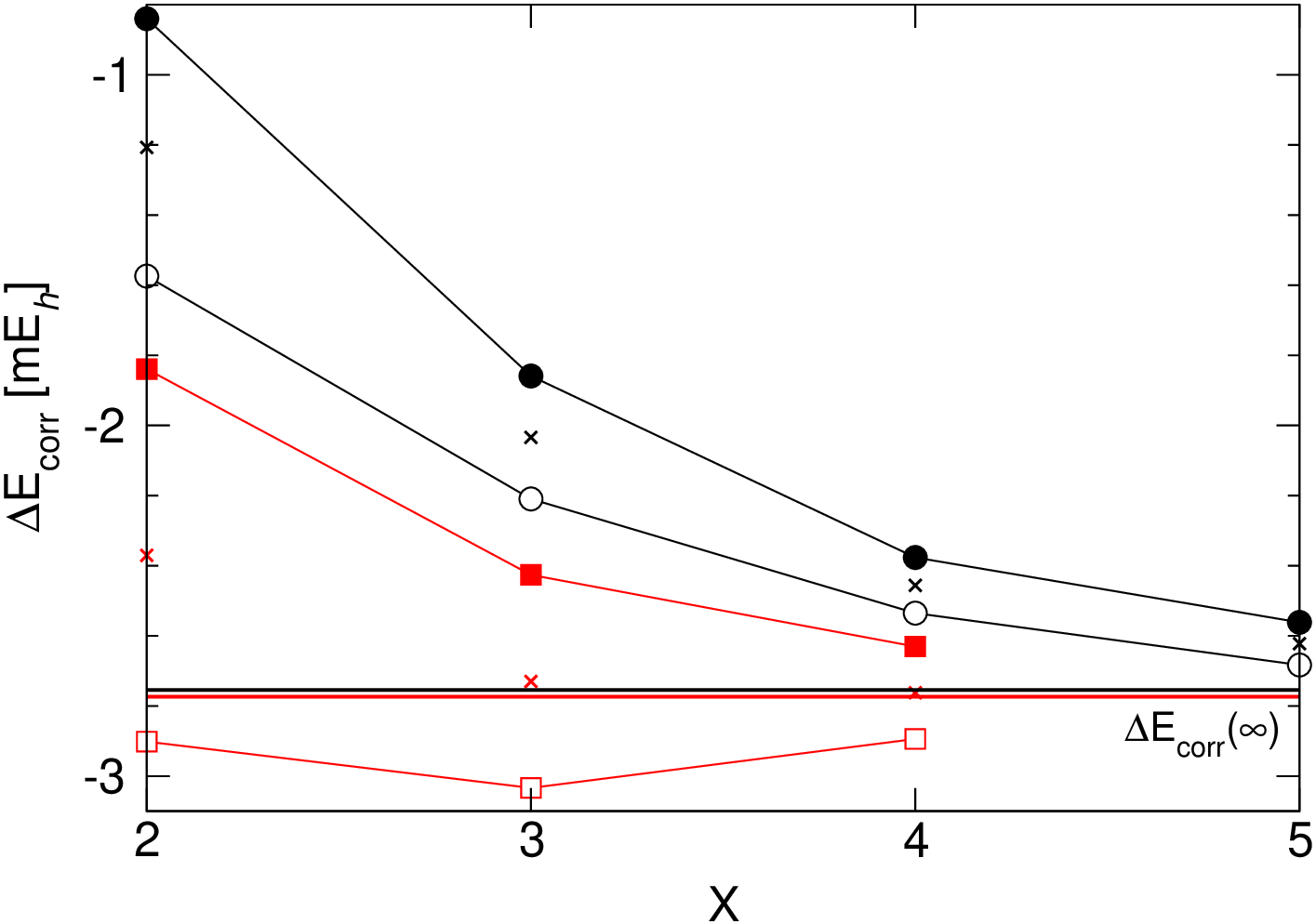}
  \caption{Basis set convergence of the correlation contribution to the
           binding energy per monomer~$\Delta E_{\rm corr}$ in~\HCl.
           Symbols are chosen as in Fig.~\ref{fig:HF_RHF_basis}.}
  \label{fig:HCl_basis}
\end{figure}

The correlation contributions to the binding energies per monomer
of~\HF{} and \HCl{}
are plotted in Figs.~\ref{fig:HF_basis} and~\ref{fig:HCl_basis}.
For each series of correlation consistent basis sets, there is
an upper curve for the CP~corrected binding energies and an lower
curve giving the bare binding energies without CP~correction
of the monomer energies. We note that the CP~corrected energies
drop monotonically with improving basis set quality but the
corresponding bare energies behave unsystematically.
The binding energy curves of~\HF{} in Fig.~\ref{fig:HF_basis} for
the aug-cc-pVXZ and d-aug-cc-pVXZ basis sets are essentially parallel,
if counterpoise correction is applied, but the associated curves for
the bare binding energies differ considerably, an effect
which is exclusively caused by the behaviour of the monomer energies
employed for the bare and CP~corrected curves.
The small difference between the two CP~corrected curves indicates
the small improvement of the description of the ground state of a
HF monomer and \HF{} by the second set of diffuse
functions in d-aug-cc-pVXZ basis sets.

The error of the correlation contribution to the binding energy,
introduced by the finite basis sets, can be estimated as the differences
of the bare and the CP~corrected values corresponding to the same
basis set series in Figs.~\ref{fig:HF_basis} and \ref{fig:HCl_basis}.
Using the cc-pV5Z binding energies, the deviation of the bare binding
energy from the CP~corrected binding energy is
%
%
$\approx 20 \%$ for~\HF{} and
%
%
$\approx 5 \%$ for~\HCl{}, the error being considerably larger for~\HF{}
compared to~\HCl{}.
The basis set error is far too large for a definitive value of
the correlation contribution to the binding energy of the infinite chains.

\begin{table}
  \caption{Basis set extrapolation of the CP~corrected correlation
           contribution to the binding energy per monomer~$\Delta E_{\rm corr}$
           of~\HF{} by means of Eqs.~(\ref{eq:atomicX3})
           and (\ref{eq:atomicXgamma}). All data are given
           in millihartree.}
  \centering
  \begin{tabular}{rcc}
    \hline
    X--Y & $\displaystyle \Delta E_{\rm corr}(\infty) \atop
            \textrm{    cc-pVXZ}^{\vphantom{/}}$
         & $\displaystyle \Delta E_{\rm corr}(\infty) \atop
            \textrm{aug-cc-pVXZ}^{\vphantom{/}}$ \\
    \hline
       $X^{-3}$ D--T & -1.037 & -1.337 \\
                D--Q & -1.433 & -1.547 \\
                T--Q & -1.637 & -1.655 \\
                Q--5 & -1.606 &        \\
    Park~\etal\ D--T & -1.215 & -1.475 \\
                D--Q & -1.496 & -1.593 \\
                T--Q & -1.620 & -1.646 \\
                Q--5 & -1.585 &        \\
    \hline
  \end{tabular}
  \label{tab:CPScorrHF}
\end{table}

\begin{table}
  \caption{Basis set extrapolation of the CP~corrected correlation
           contribution to the binding energy per monomer~$\Delta E_{\rm corr}$
           of~\HCl{} by means of Eqs.~(\ref{eq:atomicX3})
           and (\ref{eq:atomicXgamma}). All data are given in millihartree.}
  \centering
  \begin{tabular}{rcc}
    \hline
    X--Y & $\displaystyle \Delta E_{\rm corr}(\infty) \atop
            \textrm{    cc-pVXZ}^{\vphantom{/}}$
         & $\displaystyle \Delta E_{\rm corr}(\infty) \atop
            \textrm{aug-cc-pVXZ}^{\vphantom{/}}$ \\
    \hline
       $X^{-3}$ D--T & -2.300 & -2.681 \\
                D--Q & -2.604 & -2.742 \\
                T--Q & -2.760 & -2.773 \\
                Q--5 & -2.747 &           \\
    Park~\etal\ D--T & -2.409 & -2.742 \\
                D--Q & -2.642 & -2.764 \\
                T--Q & -2.750 & -2.774 \\
                Q--5 & -2.762 &           \\
    \hline
   \end{tabular}
  \label{tab:CPScorrHCl}
\end{table}

To get the basis set limit correlation contribution to the binding
energies, firstly, we apply a two-point fit based on Eq.~(\ref{eq:atomicX3}),
involving the correlation energies for the infinite chains, obtained with two
basis sets~$X$ and $Y$, and the CP~corrected ones for
the respective monomers~\cite{Halkier:BS-98}.
Secondly, Eq.~(\ref{eq:atomicXgamma}) is harnessed,
where the correlation energy of the monomer is required in a very
good approximation. As there
are no suitable R12~data for~HF and HCl, we determine the
correlation energy of the monomers by
$X^{-3}$~extrapolation~(\ref{eq:atomicX3}). For~HF, we extrapolate
the correlation energies of a cc-pV5Z and a cc-pV6Z calculation,
where basis set extension~\cite{Duijneveldt:CP-94}, \ie~the
basis set improvement due to the basis functions of neighbouring
monomers, is accounted for by surrounding the HF~monomer by the basis sets
of four additional monomers which yields~$E_{\rm corr,\ HF} = -0.314530 \Hartree$.
For the HCl~monomer, the basis sets of only two neighbouring
HCl~monomers is utilised in the extrapolation of aug-cc-pV5Z
and aug-cc-pV6Z correlation energies, yielding~$E_{\rm corr,\ HCl}
= -0.254373 \Hartree$.

Two-point extrapolations for~\HF{} and \HCl{} are displayed in
Tabs.~\ref{tab:CPScorrHF} and \ref{tab:CPScorrHCl}, and are
found to converge rapidly towards a limit with increasing
quality of the involved basis sets. The most reliable extrapolations
to the limit are obtained, if basis sets of largest~$X$
and $X+1$ are employed. For several molecules, this fit
was shown to yield the best approximation to the basis set limit,
with respect to accurate R12~data, for the correlation contribution
to the binding energy~\cite{Halkier:BS-98}.
The extrapolations obtained with cc-pVXZ
and aug-cc-pVXZ series are found to approach each other
as they should. Inspecting Figs.~\ref{fig:HF_basis} and
\ref{fig:HCl_basis}, we observe that the extrapolations
involving the best basis sets, \ie~with largest~$X$ and $X+1$,
lie well in the range where they are expected to be, leading
to the conclusion that the basis set limit is nearly reached.
The very good agreement
of the best extrapolated values [Q--5 for cc-pVXZ and T--Q for
aug-cc-pVXZ] indicates reliability for the correlation contribution
to the binding energies of both chains. Especially $X^{-3}$~extrapolations
involving double~$\zeta$ quality basis sets
in Tabs.~\ref{tab:CPScorrHF} and \ref{tab:CPScorrHCl}
are found to be less accurate than corresponding extrapolations
by Park~\etal{} with respect to the values of Q--5~extrapolation
[cc-pVXZ] or the ones of T--Q~extrapolation [aug-cc-pVXZ], as the
latter extrapolation method is independent of the convergence properties of
the series~\cite{Hill:RC-85,Helgaker:BS-97} underlying
$X^{-3}$~extrapolation~(\ref{eq:atomicX3series}).
The mean of corresponding bare and CP~corrected energies~\cite{Halkier:BS-99}
does only provide a decent approximation to the basis set limit
binding energies
for~\HF{} [Fig.~\ref{fig:HF_basis}], the results for~\HCl{} are
contradictory [Fig.~\ref{fig:HCl_basis}].

\section{Conclusion}

\begin{table}
  \caption{Basis set extrapolated binding energies per
           monomer~$\Delta E(\infty)$ of~\HF{} and \HCl{} and their
           decomposition into basis set extrapolated
           Hartree-Fock~$\Delta E_{\rm SCF}(\infty)$ and
           electron correlation~$\Delta E_{\rm corr}(\infty)$
           contributions. $\Delta E_{\rm SCF}(\infty)$~is the mean of
           the four extrapolated energies for each infinite chain, respectively, in
           Tab.~\ref{tab:RHFextra} and $\Delta E_{\rm corr}(\infty)$~is the mean
           of the two Q--5 (cc-pVXZ) and the two T--Q (aug-cc-pVXZ)
           extrapolated energies for each infinite chain, respectively,
           in Tabs.~\ref{tab:CPScorrHF} and \ref{tab:CPScorrHCl}.
           All data are given in electronvolt.}
%
%
%
%
%
%
%
%
  \centering
  \begin{tabular}{rcc}
    \hline
                 & \HF{} & \HCl{} \\
    \hline
    $\Delta E_{\rm SCF}(\infty)$  & -0.2776 & -0.01594 \\
    $\Delta E_{\rm corr}(\infty)$ & -0.0442 & -0.07521 \\
    $\Delta E(\infty)$            & -0.3218 & -0.09115 \\
    \hline
  \end{tabular}
  \label{tab:bindtotal}
\end{table}

In our study we focus on the basis set convergence of the Hartree-Fock
and the correlation energy in the \emph{infinite} chains~\HF{}
and \HCl{}. Especially in hydrogen bonded systems, the binding energy
per monomer is very small and very accurate calculations are required.
Our data shows that the error of the binding energy, employing
the cc-pV5Z basis set, leads to a deviation of the bare binding
energy per monomer of~\HF{} from the CP~corrected one by~$\approx 20\%$
and by~$\approx 5\%$ for~\HCl{} which is too large to be satisfactory.

To reduce the error, we extrapolate Hartree-Fock~\cite{Jensen:BS-99} and
correlation energies~\cite{Helgaker:BS-97,Halkier:BS-98,Halkier:BS-99,Klopper:R12-00,Park:BS-01,Huh:BS-03} where two different extrapolation schemes are studied for the
correlation energy.
The accuracy of the resulting correlation contribution to the binding
energies in Tab.~\ref{tab:bindtotal} is estimated
by the deviation of the extrapolated
binding energies for the aug-cc-pVXZ series from the ones for the
cc-pVXZ series, where the largest basis sets of the respective series
[T--Q for aug-cc-pVXZ and Q--5 for cc-pVXZ] are utilised as they
could be shown to yield the best results. The deviation is
%
%
$\approx 3\%$ or $\approx 4\%$ for~\HF{} and
%
%
$\approx 1\%$ or $\approx 0.4\%$ for~\HCl{} depending on the
extrapolation method employed, \ie~$X^{-3}$~\cite{Helgaker:BS-97,Halkier:BS-98,Halkier:BS-99,Klopper:R12-00}
or Park~\etal~\cite{Park:BS-01,Huh:BS-03}.

We would like to point out further, that Eqs.~(\ref{eq:atomicX3}) and
(\ref{eq:atomicXgamma}) facilitate to extrapolate the individual energy
increments occurring in the decomposition of the correlation
energy in terms of the incremental
scheme~\cite{Stoll:CS-92,Stoll:CD-92,Stoll:CG-92,Buth:BE-04}, separately.
This is advantageous for treating, \eg, hydrogen bonds in larger molecules
and facilitates a more accurate treatment of different atoms or fragments
in crystals.

\begin{ack}
We are very grateful to Uwe Birkenheuer, Peter Fulde and Hermann Stoll
for helpful discussions and a critical reading of the manuscript.
\end{ack}

\end{document}